# Understanding the Nonlinear Behavior and Frequency Stability of a Grid-synchronized VSC Under Grid Voltage Dips

Chen Zhang, Marta Molinas, *Member, IEEE*, Xu Cai and Atle Rygg

*Abstract*—Transients of a grid-synchronized voltage source converter (VSC) are closely related to over- currents and voltages occurred under large disturbances (e.g. a grid fault). Previous analysis in evaluating these transients usually neglect the nonlinear control effects of a VSC (e.g. phase-locked-loop, PLL). Therefore, potential stability issues related with nonlinear dynamics cannot be revealed properly. This work aims to move further in this respect. To better analyze and gain more insights into the nonlinear properties, dynamical analysis of a grid-tied VSC is conducted by parts. Specifically, the nonlinear behaviors of VSC power control loop (PCL) are firstly analyzed, in which the dynamics of PLL are assumed steady. Then, the nonlinear behaviors of PLL-dominant dynamics are further explored in detail, where the PCL is assumed steady. In this case, frequency instability and the mechanisms behind it are revealed. At last, effects of PQ controller regulation as well as controller bandwidth on the frequency stability are discussed. All the analysis and conclusions are verified by time domain simulations in PSCAD/EMTDC, where a switching model of VSC is adopted.

## I. INTRODUCTION

Voltage source converters (VSCs) are widely adopted in integrating renewable energy power generations (e.g. wind and solar power) with AC grid [1], as well as in combining two asynchronous power system by high voltage dc (HVDC) technology [2].

Despite the fast and flexible power control capability of VSCs, recent experience identify that the VSC synchronized by a phase-locked-loops (PLL) is susceptible to oscillate against a weak AC grid, e.g. a case of wind farm case in [3] and photovoltaic plant in [4]. It inspires the researchers to explain and analyze these oscillation behaviors, and many endeavors are made in this respect in recent years (e.g. [5] and [6]). Due to the oscillations of VSCs usually occurred at a steady point, this allows linear-based methods to identify and predict the forward behavior around this point. Among these, the impedance-based analysis method is prevailing, due to its intuitive representation and the capability of verification by simulations or field measurements.

The impedance models of a typical three phase VSC can be different in formats according to the linearization methods they adopt. Typically, linearizing in sequence domain [7] results in a sequence impedance (e.g. [8] and [9]), whereas linearizing in *dq* domain [10] yields a *dq* impedance (e.g. [11] and [12]). Recently, other modeling methods, e.g. a phasor-based [13], a modified sequence domain based [14], a single-input and single output based [15] and a complex transfer function based [16] methods are available, in which the properties of converters can be explained more intuitively, e.g. the mirror coupling effects [14] originated from the *dq* asymmetry of impedance matrices [17]. Despite different modeling techniques, all these models are capable for identifying oscillation behaviors in a grid-tied VSC system by applying Nyquist-based stability analysis [18]. A useful finding is that, PLL is of great importance for VSC small signal stability, in particular with a weak AC grid condition (e.g. [11] and [19]). Moreover, behaviors of PLL can be physically interpreted as a current-controlled-voltage-source (CCVS) in an equivalent RLC circuit of VSC-grid system, where the CCVS can exhibit negative damping to the equivalent RLC circuit if conditions in [20] are met.

However, the impedance-based methods are in linear domain which can only predict the dynamics around a steady point, if the oscillation diverges, the subsequent behavior can no longer be predicted accurately, e.g. a limit cycle. In [21], a large signal impedance model is proposed to study the sustained oscillations, where the PWM saturation is properly modeled compared to the typical impedance model, e.g. [11]. In [22] and [23], a state space nonlinear model of VSC is proposed, where the bifurcation phenomenon is observed. However, these analysis focus on the hard nonlinearities (e.g. saturation), whereas the dynamical nonlinearities (e.g. PLL) are simplified.

On the other hand, a phasor-based analysis of a wind farm with grid voltage dips are studied in [24] and [25], where the wind farm model is simplified to a current source.

"This work is rejected due to some of the explanations are not clear. And experimental validation is lacking particularly when analyzing PCL controller behavior".

Chen Zhang, Department of Electrical Engineering, Shanghai Jiao Tong University, Shanghai, China (email: nealbc@sjtu.edu.cn).

Marta Molinas, Department of Engineering Cybernetics, Norwegian University of Science and Technology, Trondheim, Norway (marta.molinas@ntnu.no).

Xu Cai, Department of Electrical Engineering, Shanghai Jiao Tong University, Shanghai, China (email: xucai@sjtu.edu.cn).

Atle Rygg, Department of Engineering Cybernetics, Norwegian University of Science and Technology, Trondheim, Norway (email: atle.rygg@ntnu.no).

Requirements on the limitation of active current injection, in the presence of grid voltage dips are emphasized. This methodology is very similar to the power limit analysis of a synchronous generator (SG) and infinite bus-bar in conventional power system. In [26], a similar conclusion was derived from the perspective of positive feedback effects of PLL [24] in a VSC system.

Although these works ([24]-[26]) lack dynamical analysis of nonlinearities, they are helpful and illuminated in reconsideration of VSC transients that we thought mostly owing to the circuit transients, e.g. over-currents. Worth to mention that a recent work in [27] analyzed the transient stability issue of a power synchronization controlled (PSC, proposed in [28]) VSC. Due to the dynamic model of PSC-VSC is very similar to the SG intrinsically, so that SG-based theory can be applied directly.

However, in this paper VSC with typical control method (i.e. vector-orientation-based control) will be analyzed. To the authors' knowledge, research is lacking in this respect. The rest of the paper is organized as bellows: in section. II, the nonlinear behavior of power control loop (PCL) is analyzed firstly, where the PLL is assumed steady. Then, the PLL- dominant nonlinear behavior is exploited in section. III, in which the frequency instability is identified and mechanisms behind it are revealed in detail. Based on the developed knowledge, impacts of PCL on the frequency stability are further discussed in section IV. Finally, section V draws the main conclusion. All the simulations are conducted in PSCAD/EMTDC, where a switching model of grid-tied VSC system is adopted.

## II. ANALYSIS OF POWER CONTROL LOOP DOMINANT NONLINEAR DYNAMICS

### A. Description of study system

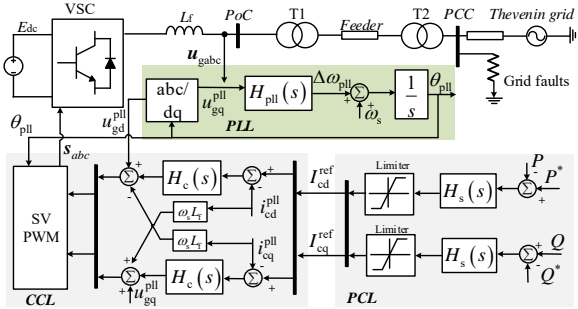

Fig. 1 Schematic of a typical grid connected VSC system

A typical grid-synchronized VSC is shown in Fig. 1, it has a current control loop (CCL), a PCL and a PLL ($H_c$, $H_s$, $H_{pll}$ are PI controllers for CCL, PCL and PLL respectively). Typically, the VSC is connected to the bulk grid via two step-up transformers (T1 and T2) to boost voltage to the transmission level, hence the "equivalent grid" seen from the VSC (i.e. at the point of connection, PoC) is relatively weak, even though the impedance of the Thevenin grid (i.e. impedance seen from the point of common coupling, PCC) can be very small. This weak grid condition can lead to oscillation if the PLL and CCL are not decoupled [20] in time-scale (or bandwidth in frequency domain), this small-signal-stability issue has been extensively studied in literatures as discussed before. This work therefore assumes a properly designed CCL, which means CCL is much faster than PLL or PCL. Consequently, CCL can be assumed steady in PLL or PCL time-scale.

Table I MAIN PARAMETERS

| Symbol | Name | Value |
|---|---|---|
| $S_n$ | rating/base | 2 MW |
| $U_n$ | nominal/base voltage | 690 V |
| $V_{dc}$ | dc voltage | 1200 V |
| $f_{sw}$ | switching frequency | 2.4 kHz |
| $f_s$ | fundamental frequency | 50 Hz |
| $L_f$ | filter inductance | 7.6 e-5 H |
| $L_T$ | Leakage inductance of $T_1$ and $T_2$ | 0.1 p.u. |
| $H_{pll}$ | PLL controller | $k_p^{pll} + k_i^{pll}/s$ |
| $H_s$ | PQ controller | $k_p^s + k_i^s/s$ |

### B. Modeling of PCL dominant dynamics

Model reduction is necessary to render an analytical study of nonlinear behaviors in a complex system. This section aims to analyze PCL dominant dynamics, hence PLL is assumed steady temporarily. On the basis of this assumption, the voltage equation from PoC ($U_g^{pll}$) to the Thevenin grid ($U_s^{pll}$) is:

$$U_g^{pll} = j\omega_{pll}L_\Sigma I_c^{pll} + U_s^{pll} \quad (1)$$

where, the output currents ($I_c^{pll}$) of VSC are assumed steady due to their fast dynamics, i.e. $I_c^{pll} \approx I_c^{ref}$. All the vectors are projected to the PLL reference frame (denoted by superscript 'pll'). Since the PLL is assumed steady in this case, it has $\omega_{pll} = \omega_s$. In PLL fame, the Thevenin grid voltage can be represented as: $U_s^{pll} = U_s \cos\theta_0 - jU_s \sin\theta_0$, where $\theta_0$ is the steady angle difference between $U_g$ and $U_s$. $L_\Sigma$ is the lumped system inductance seen from PoC, it can be quantified by the short circuit ratio (SCR), i.e. in per unit format $\bar{L}_\Sigma = 1/SCR$.

The output power of VSC in complex format is:

$$P + jQ = U_{pcc}^{pll}(I_c^{pll})^* = j\omega_{pll}L_\Sigma|I_c^{pll}|^2 + U_s^{pll}(I_c^{pll})^* \rightarrow$$

$$\begin{cases} P = U_s\cos\theta_0 \cdot I_{cd} - U_s\sin\theta_0 \cdot I_{cq} \\ Q = \omega_s L_\Sigma(I_{cd}^2 + I_{cq}^2) - (U_s\cos\theta_0 I_{cq} + U_s\sin\theta_0 I_{cd}) \end{cases} \quad (2)$$

It can be observed from (2) that the active power (P) is a linear combination of dq currents, whereas Q is not.

Next, the dynamical equation of PCL can be further derived from the control blocks in Fig. 1 as:

$$\begin{cases} \dfrac{dx_d}{dt} = k_i(P^* - P) \\ \dfrac{dx_q}{dt} = k_i(Q^* - Q) \\ I_{cd} = x_d + k_p(P^* - P) \\ I_{cq} = -x_q - k_p(Q^* - Q) \end{cases} \quad (3)$$

In which, $x_d, x_q$ are the output states of PI integrators, while $k_p$ and $k_i$ are PI parameters of $H_s$.

Consequently, (2) and (3) consist of the model of PCL-dominant dynamics.

*C. Nonlinear behavior analysis*

Due to the developed PCL model is of second-order, phase portraits can therefore be adopted to study nonlinear its behaviors. In Fig. 2, current phase portraits with varying PQ controller ($H_s$) parameters are plotted.

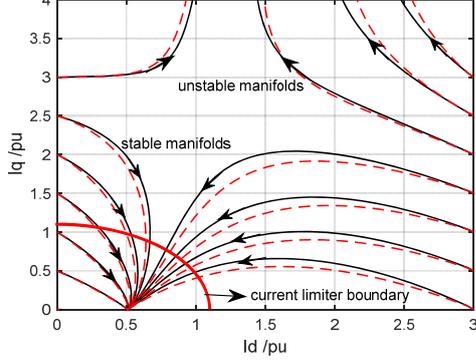
(a) Phase portraits with varying $k_P$ (solid line $k_P = 0.1$, dotted line $k_P = 0.2$)

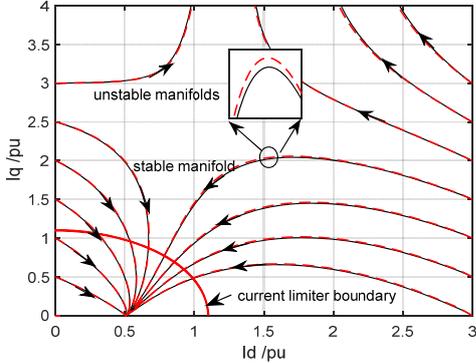
(b) Phase portraits with varying $k_i$ (solid line $k_i = 20$, dotted line $k_i = 40$)

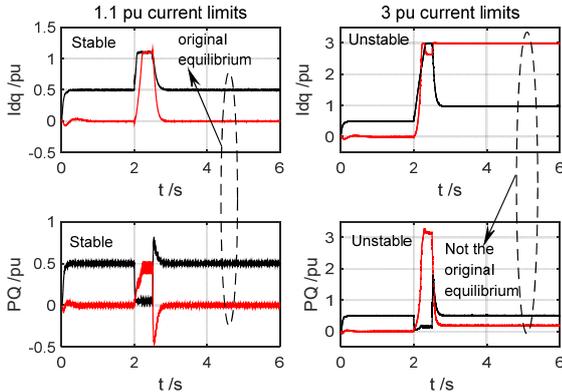
(c) Time domain simulations (black lines denote $I_d/P$, red lines denote $I_q/Q$)
Fig. 2 Nonlinear behavior analysis of PCL dominant dynamics (In steady, P = 0.5 pu, Q = 0 pu, SCR = 4. In (c), $k_p = 0.1$, $k_i = 20$)

From Fig. 2 (a) we can observe that, the states ($I_{cd}$, $I_{cq}$) can be either attracted by the stable manifolds or driven away by the unstable manifolds, dependent on their initial values. In other words, the equilibrium (0.5, 0) in this case has a limited region of attraction, consequently, any initial states that are closely enough to the equilibrium can be attracted, otherwise a divergent of states can occur. On the other side, comparing the dotted line and solid line we can identify that, increasing the proportional gain of *PQ* controller ($H_s$) has a negative impacts on stability (i.e. region of attraction becomes smaller), whereas the integral gain has negligible effects on the phase portraits as depicted in Fig. 2 (b).

To validate the analysis, time domain simulations are presented in Fig. 2 (c), where the transient responses of currents ($I_{dq}$) and active/reactive power (*P/Q*) under a symmetrical grid fault are measured. In accordance with the model assumption, PLL frequency is locked to 1.0 p.u. after VSC is synchronized (i.e. at 1.8s). As shown in Fig. 2 (c), if the currents are limited to a relatively small value (left columns), the system can converge to the original equilibrium (i.e. ($I_d$, $I_q$) = (0.5, 0)) after grid fault is cleared, this case is consistent with the analysis in Fig. 2 (a) with initial states within the region of attraction.

However, if the currents are limited to a relatively large value (right column in Fig. 2 (c)), the system cannot converge to its original equilibrium but to another "steady" point. It should be noted that, this "steady" point is not an equilibrium because the reactive power is not converge to its set-point, which is *Q* = 0. Hence, a condition: $\dot{x}_d = 0, \dot{x}_q \neq 0$ can be obtained from (3). Further based on (2) we can notice that, it has *three* unknowns (i.e. $Q$, $I_d$, $I_q$) but only with *two* independent equations, resulting in infinite numbers of solutions, and the "steady" point shown in Fig. 2 (c) (right column) is essentially one of them. Apparently, this steady point is not an equilibrium since condition $\dot{x}_d = 0, \dot{x}_q = 0$ is not satisfied.

Another finding should be addressed is that, although the theoretical analysis can be unstable, this maybe unachievable due to the current limits in practice should be small for protection purpose (e.g. typical values are $I_{max} = 1.1 I_n$). Under this circumstances, all the initial states of currents can be constrained in the region of attraction (red arcs in Fig. 2 (a) and (b)). Consequently, PCL-dominant dynamics can be considered as *absolute* stable in this case.

III. ANALYSIS OF PLL DOMINANT NONLINEAR DYNAMICS

Previous section analyzed the PCL dominant nonlinear behaviors, in which PLL is assumed steady. In this section, the nonlinear behavior of PLL-dominant dynamics will be analyzed in detail.

*A. Modeling of PLL dominant dynamics*

To effectively model and analyze the PLL-dominant dynamics, a constant current reference is assumed, which means regulation of *PQ* controller is not considered in this section, but it will be discussed in detail in section IV.

Again, due to the fast current dynamics, in PLL time-scale it has: $\boldsymbol{I}_c^{pll} \approx \boldsymbol{I}_c^{ref}$. In this case, the input of PLL is the *q* axis voltage at PoC, i.e. $u_{gq}^{pll} = \text{Im}\{\boldsymbol{u}_g^{pll}\}$ from (1). In combination with PLL control blocks in Fig. 1, the following equations can be obtained:

$$\begin{cases} \dfrac{d\Delta\omega_{pll}}{dt} = k_p^{pll} \cdot \dot{u}_{gq}^{pll} + k_i^{pll} u_{gq}^{pll} \\ \dfrac{d\delta_{pll}}{dt} = \Delta\omega_{pll} \end{cases} \quad (4)$$

In which, $u_{gq}^{pll} = \omega_{pll} L_\Sigma I_{cd}^{ref} - U_s \sin\delta_{pll}$, $\delta_{pll} = \theta_{pll} - \theta_s$. $k_p^{pll}$ and $k_i^{pll}$ are the PI parameters of $H_{pll}$. In the later analysis, PLL bandwidth ($\alpha_{pll}$) is frequently used instead of PI parameters, in which $k_p^{pll} = 2\alpha_{pll}/U_s$ and $k_i^{pll} = 2\alpha_{pll}^2/U_s$ are adopted [20].

Converting (4) into per unit format, yields:

$$\begin{cases} T_{pll}\dfrac{d\Delta\bar{\omega}_{pll}}{dt} = -D_{pll}\Delta\bar{\omega}_{pll} + \bar{T}_m - \bar{T}_e \\ \dfrac{d\delta_{pll}}{dt} = \omega_b \Delta\bar{\omega}_{pll} \\ D_{pll}(\delta_{pll}) = \dfrac{k_p^{pll}}{k_i^{pll}}\left(\bar{U}_s \cos\delta_{pll}\omega_b - \bar{L}_\Sigma \bar{I}_{cd}^{ref}\right) \\ T_{pll} = \dfrac{\omega_b - k_p^{pll}\bar{L}_\Sigma \bar{I}_{cd}^{ref}}{k_i^{pll}} \\ \bar{T}_m = \bar{\omega}_s \bar{L}_\Sigma \bar{I}_{cd}^{ref}, \bar{T}_e = \bar{U}_s \sin\delta_{pll} \end{cases} \quad (5)$$

where $\omega_b$ is the base valued angular frequency. $T_{pll}$ is a constant, whereas $D_{pll}$ is $\delta_{pll}$ dependent. For a small value of $\delta_{pll}$, $D_{pll}(\delta_{pll})$ is positive; otherwise it can be negative. $\bar{T}_m$ is a constant input, $\bar{T}_e$ is the nonlinear state feedback.

Consequently, the nonlinear model of PLL-dominant dynamics is developed in (5). One may already identified that this model resembles the motion equation of a SG, which means SG-based analysis methods are applicable. To further address these similarities, the notation for a SG is adopted to the definition of PLL variables, e.g. $\bar{T}_m$, $\bar{T}_e$, although in physical viewpoint $\bar{T}_m$ and $\bar{T}_e$ in this model are essentially voltages.

### B. Nonlinear behavior analysis

Likewise, (5) is a second-order system rendering the phase portrait analysis of its nonlinear behaviors, where $(\Delta\bar{\omega}_{pll},\delta_{pll})$ are the two state variables.

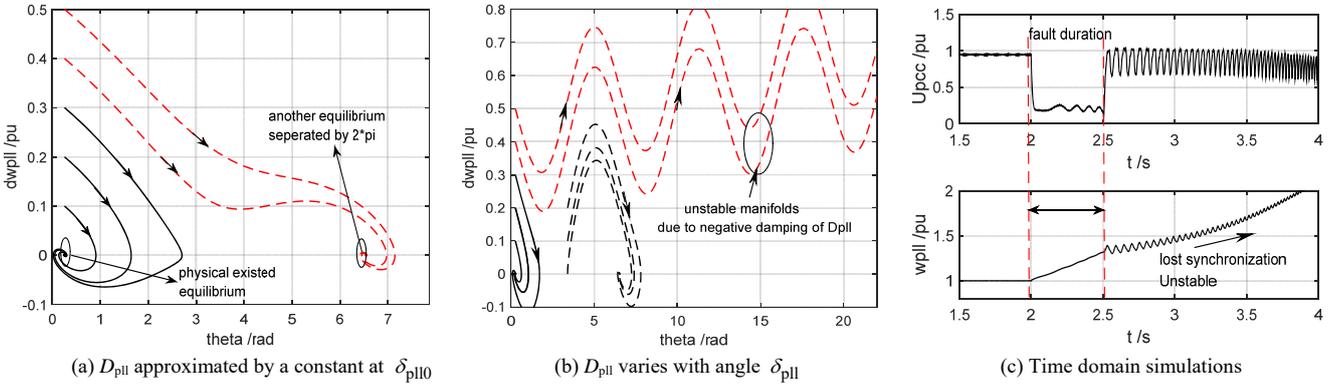

(a) $D_{pll}$ approximated by a constant at $\delta_{pll0}$    (b) $D_{pll}$ varies with angle $\delta_{pll}$    (c) Time domain simulations

Fig. 3 Nonlinear behavior analysis of PLL dominant dynamics (PLL bandwidth is 20 Hz, i.e. $k_p^{pll}=20$, $k_i^{pll}=800$, SCR = 4, $\bar{I}_{cd}^{ref}=1$ pu. In (c), a symmetrical grid fault is applied at PCC with a duration of 500ms.)

Firstly in Fig. 3 (a), the phase portraits of (5) with varying initial states $(\Delta\omega_{pll0},\delta_{pll0})$ are plotted, where the damping term is approximated by a positive constant $D_{pll}(\delta_{pll0})$ (i.e. evaluated at $\delta_{pll0}$). We can observe that for a small initial states around the original equilibrium, states can converge to this point after several cycles' motion. However, if the initial states are far away from this point, they can converge as well, but to another equilibrium that separated by $2\pi$. This is an intrinsic property of nonlinear system (5) that, there is an equilibrium set (i.e. $\delta_{pll0} + 2k\pi$, $k = \pm1, \pm2...$) that is locally stable, and states can always converge to the equilibrium set due to $D_{pll}(\delta_{pll0})$.

However, in fact we know that the damping term $D_{pll}$ a nonlinear function of $\delta_{pll}$, which means it can be negative if $\delta_{pll}$ is large (e.g. $\delta_{pll} \in \left[2k\pi + \dfrac{\pi}{2}, 2k\pi + \dfrac{3\pi}{2}\right], k = 0, \pm1...$). Therefore, a detailed phase portraits including the effects of $D_{pll}(\delta_{pll})$ is plotted in Fig. 3 (b). Clearly, the system can be unstable (dotted line marked with red color) if the initial states of $(\Delta\omega_{pll},\delta_{pll})$ are large. This feature cannot be captured by Fig. 3 (a) since the approximated constant $D_{pll}(\delta_{pll0})$ is positive.

A time domain analysis is further presented in Fig. 3 (c), where the PLL frequency and the magnitude of PCC voltage are measured. In order to have a large deviation of initial states, a symmetrical grid fault is applied at PCC with a duration of 500ms, we can clearly identify that the PLL lost synchronization with grid after grid fault is cleared. Moreover, it exhibits a similar manner as the phase portrait in Fig. 3 (b).

Mechanisms behind this frequency instability will be explored in the subsequently.

### C. Mechanism of frequency instability

The Equal Area Principle (EAP) developed for a SG [29] is applicable to the analysis of frequency stability, due to their similarities in dynamical equations.

According to (5), characteristics of $\bar{T}_m = \bar{\omega}_s \bar{L}_\Sigma \bar{I}_{cd}^{ref}$ and $\bar{T}_e = \bar{U}_s \sin\delta_{pll}$ can be illustrated by curves in Fig. 4 (a). It can be observed that a grid fault can change the characteristic of $\bar{T}_e$ abruptly, whereas $\bar{T}_m$ remains constant (i.e. $\bar{T}_m^{0+} = \bar{T}_m^{0-}$) due to the current is assumed steady. The magnitude differences between $\bar{T}_e^{0+}$ and $\bar{T}_m^{0+}$ can result in the acceleration of system states: $(\Delta\omega_{pll}, \delta_{pll})$. Then, after a short period, the grid fault is cleared (e.g. at point C), where characteristic curve of $\bar{T}_e^{0+}$ changes back to its pre-fault value ($\bar{T}_e^{0-}$), so that $\Delta\omega_{pll}$ starts decelerating due to $\bar{T}_e^{0-}$ is greater than $\bar{T}_m^{0-}$. However, $\delta_{pll}$ will remain increasing until $\Delta\omega_{pll}$ becomes zero again.

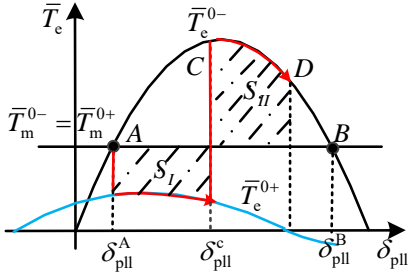

(a) Mechanism analysis (Superscript "0-" and "0+" denote pre-fault and post-fault; point A is pre-fault equilibrium, point C is fault clearing point)

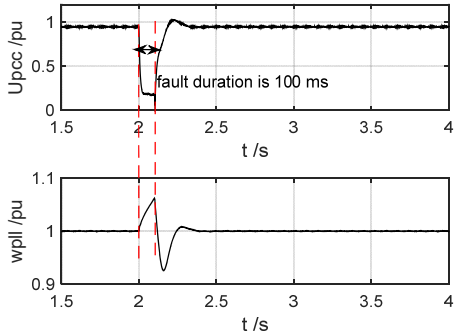

(b) Time domain simulations (conditions are the same as Fig. 3 (c), except that fault duration is reduced to 100ms)

Fig. 4 EAP-based mechanism analysis of frequency stability

The EAP claims that if $\exists S_{II}(\Delta\omega_{pll} = 0, \delta_{pll} \leq \delta_{pll}^B): S_I = S_{II}$, then the system is referred to as *first swing stable*. It should be noted that, although the nonlinear system (5) can converge to an equilibrium set (e.g. in Fig. 3 (a)) as discussed before, only the principal one i.e., $\Delta\omega_{pll} = 0, \delta_{pll} = \delta_{pll0}$ is of interests or physical significance. Because the transition from one equilibrium to another can exhibit large transients in currents or voltages, which are not allowed due to limited stress of physical components, e.g. VSC. For a SG, this also indicates a "pole slip" operation, which can be detrimental.

From EAP we can also find that, if the fault clearing angle $\delta_{pll}^C$ is small (i.e. fault is cleared fast), then there is more margin for the deceleration area that stability can be assured. To illustrate this character, a time domain simulation is shown in Fig. 4 (b), where the fault is cleared faster compared to the one in Fig. 3 (c). Clearly, the frequency of PLL in this case is stable, proving that a small $\delta_{pll}^C$ is helpful for stability.

Hence, the fault clearing angle can be a measure of frequency stability margin, question is how to evaluate it. Regarding this, the critical condition that ensuring a first swing stable system is of most interest, i.e. there exists a critical angle that EAP is met, i.e. $\exists \delta_{pll}^{CCA}: S_I = S_{II}^{max}(\Delta\omega_{pll} = 0, \delta_{pll}^{max} = \delta_{pll}^B)$. This angle is referred to as Critical Clearing Angle (CCA).

### D. Analysis of frequency stability margin

The CCA can be calculated by numerical method if analytical model of $\bar{T}_e^{0+}$, $\bar{T}_e^{0-}$ and $\bar{T}_m$ are known, in which $\bar{T}_e^{0-}$ and $\bar{T}_m$ can be obtained from (5), whereas $\bar{T}_e^{0+}$ is fault-dependent.

To calculate $\bar{T}_e^{0+}$, a circuit analysis of grid fault is necessary. In accordance with PLL modeling, the equivalent circuit of Fig.1 can be drawn in Fig. 5 (a), where $Z_{\Sigma T}$ is the lumped line impedance seen from PoC, $Z_S$ is the source impedance seen from PCC, $Z_f$ is the short circuit impedance applied at PCC.

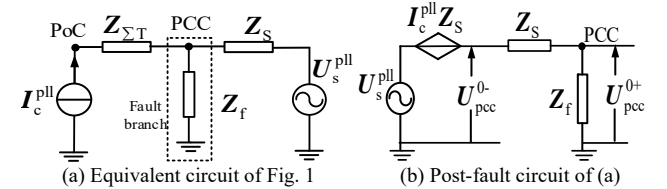

(a) Equivalent circuit of Fig. 1    (b) Post-fault circuit of (a)

Fig. 5 Equivalent circuit under a symmetrical grid fault at PCC

A Thevenin equivalent circuit seen from the fault branch can be further developed in Fig. 5 (b), in which the post-fault PCC voltage can be calculated as: $U_{pcc}^{0+} = k_f \cdot U_{pcc}^{0-}$, where $U_{pcc}^{0-} = U_s^{pll} + I_c^{pll} Z_s$ and $k_f = Z_f / (Z_f + Z_s) = k_f e^{j\varphi_f}$.

Therefore, the post-fault characteristic of $\bar{T}_e^{0+}$ is obtained:

$$\bar{T}_e^{0+} = \text{Im}\{U_{pcc}^{0+}\} = k_f U_{pcc}^{0-} \sin(\delta_{pll} - \varphi_f) \quad (6)$$

where, $\delta_{pll} = \theta_{pll} - \theta_{pcc}$ is redefined, meanwhile $\bar{T}_e^{0-}$ in (5) is modified to $\bar{T}_e^{0-} = U_{pcc}^{0-} \sin\delta_{pll}$, and $\bar{T}_m$ is modified to $\bar{T}_m = \bar{\omega}_s \bar{L}_{\Sigma T} \bar{I}_{cd}^{ref} + \bar{R}_{\Sigma T} \bar{I}_{cq}^{ref} \approx \bar{\omega}_s \bar{L}_{\Sigma T} \bar{I}_{cd}^{ref}$.

From (6) we can observe that, in geometry, $\bar{T}_e^{0+}$ is essentially a curve deformation of $\bar{T}_e^{0-}$, consequently $\bar{T}_e^{0+}$ can be easily drawn in Fig. 4 (a) by shifting and compressing curve $\bar{T}_e^{0-}$. As the grid impedance $Z_s$ is mostly inductive, and if $Z_f$ is inductive as well, there exhibits no phase shift according to the expression of $k_f$ (i.e. $\varphi_f = 0$). Otherwise, for a resistive short circuit branch, the phase shift can be $\varphi_f \in \left(-\frac{\pi}{2}, 0\right)$.

Normally, effects of $\varphi_f$ is negligible since the acceleration and deceleration areas ($S_I$ and $S_{II}$) are primarily determined by the magnitude of $\bar{T}_e^{0+}$, particularly a severe grid fault in this work.

Based on the models of $\bar{T}_e^{0+}$, $\bar{T}_e^{0-}$ and $\bar{T}_m$, the CCA can be calculated numerically from the nonlinear algebraic equation:

$$S_I = S_{II}^{\max} \rightarrow$$
$$\int_{\delta_{pll}^A}^{\delta_{pll}^{CCA}} \left(\bar{T}_m - \bar{T}_e^{0+}\right) d\delta_{pll} = -\int_{\delta_{pll}^{CCA}}^{\delta_{pll}^B} \left(\bar{T}_m - \bar{T}_e^{0-}\right) d\delta_{pll} \quad (7)$$

If resubstituting the numerical results of CCA into the dynamical equation of $\Delta\omega_{pll}$ in (5), the corresponding Critical Clearing Time (CCT) can be estimated as:

$$t_{CCT} = \frac{\left(\delta_{pll}^{CCA} - \delta_{pll}^A\right)}{k_0}\sqrt{\frac{2S_I\omega_b}{T_{pll}}} \quad (8)$$

where $k_0$ is a coefficient that used to estimate the integral of $\int_0^{t_{CCT}} \omega_b \Delta\omega_{pll} dt$ by $\left(k_0\omega_b\Delta\omega_{pll}^C\right)\cdot t_{CCT}$, in this work $k_0 = 2/3$ is adopted.

CCT is more intuitive compared to the CCA, since the time can be measured conveniently. To illustrate the feasibility of CCT in evaluating stability margin, a cluster of numerical CCT curves with varying PLL bandwidth are plotted in Fig. 6 (a).

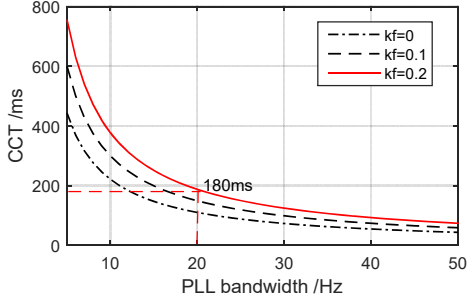

(a) CCT with varying PLL bandwidth (SCR = 4, $\bar{L}_{\Sigma T} = 0.2$ pu, $\bar{I}_{cd}^{ref} = 1$ pu, $\varphi_f = 0$)

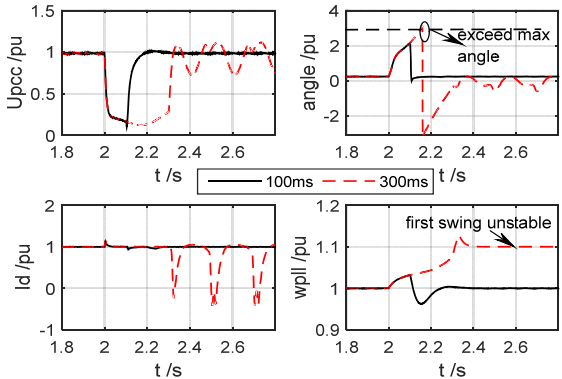

(b) Time domain simulations (a symmetrical grid fault is applied at 2s, the maximum angle denotes $\delta_{pll}^B = \pi - \delta_{pll}^A \approx 3$ rad, $\delta_{pll}^A = 0.253$ rad, a frequency limiter for PLL is at 1.1 pu)

Fig. 6 Analysis of stability margin based on CCT

It can be identified that by increasing PLL bandwidth CCTs are reducing, which means stability margin is reduced. On the other hand, the magnitudes of $\bar{T}_e^{0+}$ (i.e. voltage dips at PCC) can also affect CCT to some extends, but its effects can be limited if PLL bandwidth is large.

To further address the validity of numerical analysis, a time domain simulation is conducted in Fig. 6 (b). In which, a 20 Hz PLL bandwidth is selected, where we can find from Fig. 6 (a) that the CCT = 180ms, which means the system is stable if the fault is cleared at a duration less than this CCT value. Therefore, a100ms (<CCT) and a 300ms (>CCT) fault clearing time are compared in simulations. It is shown in Fig. 6 (b) that, for the 100ms case, the system are stable after fault is cleared (solid lines in black). However, for the 300ms case the system are unstable after grid fault is cleared (dotted line marked with red). Hence, the conclusion is consistent with numerical analysis.

Furthermore, from the current waveform (Fig. 6 (b)) we can obtain that, the current assumption made in PLL time-scale is feasible due to it can remain steady despite the varying of PLL frequency in the fault period. By exploring the angle waveform further we can further observe that, it exceeds the maximum allowed angle (i.e. $\delta_{pll}^B$ in Fig. 4 (a)) in the fault period, resulting in an unstable first swing system.

## IV. IMPACTS OF PCL ON PLL DOMINANT DYNAMICS

Nonlinear dynamics of PCL and PLL are studied separately in previous sections. In fact, PCL can affect PLL dominant dynamics by the regulation of *PQ* controller, i.e. through the current references. Objective of this section is to discuss the effects of *PQ* controller regulation on frequency stability.

### A. Impacts of PQ controller on PLL frequency dynamics

In the beginning, a vector-based analysis is conducted to see the reactions of *PQ* controller and the consequences on the frequency dynamics in the presence of grid fault.

*1) A vector-based analysis*

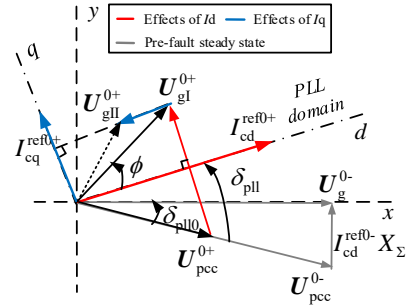

Fig. 7 Analysis of vectors motion with PCL regulation (line resistances are neglected)

Considering a voltage dip at PCC and a unity power factor operation of VSC, the pre-fault steady voltage vectors are shown in Fig. 8 (a) (gray line). As the PCC voltage drops to $U_{pcc}^{0+}$, the active power regulator will increase $\bar{I}_{cd}^{ref}$ to try to maintain a constant power, meanwhile the voltage at PoC ($U_{gI}^{0+}$) will rotate forward due to the increased $\bar{I}_{cd}^{ref0+}$, producing a phase displacement $\phi$ with respect to PLL reference frame. Since the actual current vector is assumed steady in PLL frame, this phase displacement will result in a nonzero reactive power ($Q$). As a result, the $Q$ controller will therefore take effects to produce a positive $I_{cq}^{ref0+}$, aiming to align the current vector with the newly formed $U_{gII}^{0+}$ again.

*2) Effects of P and Q control regulation*

Based on the vector analysis we can observe that, although $U_{gI}^{0+}$ rotates to $U_{gII}^{0+}$ due to the change of $I_{cq}^{ref0+}$, its projection on the $q$ axis does not change, which means $I_{cq}^{ref0+}$ will not lead to the acceleration of $\delta_{pll}$.

This finding can be obtained from the analytical model of $\bar{T}_m$ as well, where if the line resistances are neglected due to high voltage condition: $\bar{T}_m = \bar{\omega}_s \bar{L}_{\Sigma T} \bar{I}_{cd}^{ref} + \bar{R}_{\Sigma T} \bar{I}_{cq}^{ref} \approx \bar{\omega}_s \bar{L}_{\Sigma T} \bar{I}_{cd}^{ref}$, which means $\bar{T}_m$ is mostly affected by $\bar{I}_{cd}^{ref}$.

Consequently, PLL frequency dynamics are mostly affected by the regulation of $P$ controller, whereas effects of $Q$ controller can be negligible if the line resistances are small.

*3) Effects of PQ controller bandwidth*

In according with the EAP (Fig. 4 (a)) we know that the acceleration area $S_I$ can be enlarged if $\bar{T}_m$ increases. Furthermore the increasing velocity of $\bar{I}_{cd}^{ref0+}$ depends on the bandwidth of $P$ controller, if $P$ controller bandwidth is high, then $\bar{I}_{cd}^{ref0+}$ will increase quickly and even reach the current limits in a short time, resulting in a large acceleration area $S_I$ which is detrimental for frequency stability.

*B. Simulation study*

Fig. 8 (a) initially presents a comparison of transient responses of VSC with and without $Q$ controller. In can be observed that the differences in PLL frequency are negligible, proving the conclusion for $Q$ controller regulation is correct.

Further in Fig. 8 (b), the effects of controller bandwidth are evaluated. It can be observed that active and reactive currents with small $PQ$ bandwidth (10 Hz) ramp slower than that with a large $PQ$ bandwidth (20 Hz).

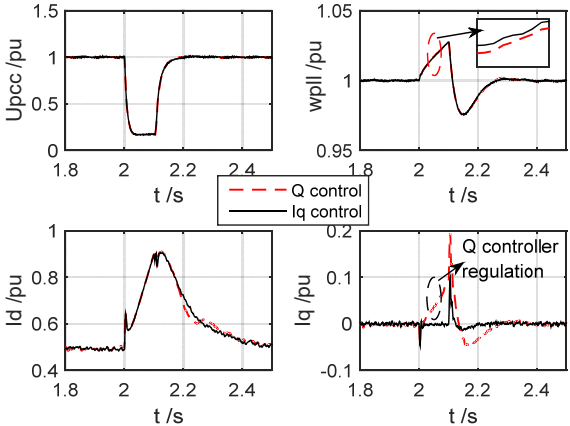

(a) Effects of PQ regulation ($P$ = 0.5 pu, $PQ$ controller bandwidth is 10 Hz. Dotted lines denote $Q$ control mode, solid lines denote $I_q^{ref} = 0$ mode)

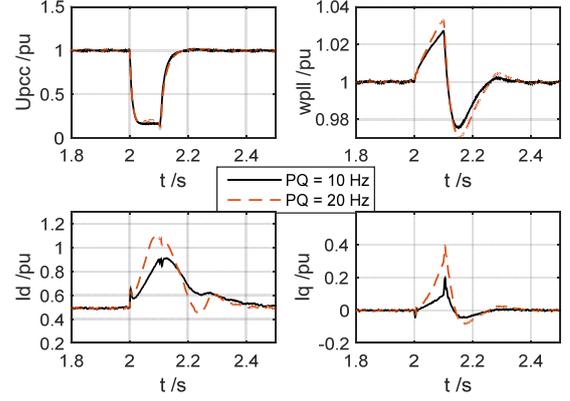

(b) Effects of PQ bandwidth ($P$ = 0.5 pu, dotted lines denote $PQ$ = 10 Hz, solid lines denote $PQ$ = 20 Hz)

Fig. 8 Simulation study of $PQ$ controller on frequency stability (a symmetric fault is applied at PCC at 2s with a duration of 100 ms, SCR = 4 (pure inductive), PLL bandwidth is 20 Hz)

As the result, the acceleration area with small $PQ$ controller bandwidth will be small either, resulting in a smaller frequency deviation as shown in Fig. 8 (b) (first graph of second column), proving the conclusion for $PQ$ bandwidth analysis is correct.

In addition, on the basis of above-mentioned analysis, it is not difficult to identify that the essential factor for frequency instability is $\bar{T}_m$, which is proportional to the active current injected into the grid. Hence, requirements on the limit of active current proposed in [24] and [25] is well supported in this work by a dynamical analysis.

## V. CONCLUSION

This paper aims to provide a groundwork for understanding the nonlinear behavior of a grid-tied VSC and its consequences on stability. To gain more insights into the properties, nonlinear behaviors of VSC control systems are studied by parts i.e. PCL and PLL respectively, then impacts of PCL on PLL dominant dynamics are further analyzed according to the established knowledge. The main findings are as follows.

1) Considering the limited stress of physical components, PCL with a steady PLL can be regarded as an absolute stable system.
2) Nonlinear dynamics of PLL with a steady PCL can exhibit frequency instability under severe grid voltage dips. Mechanisms behind it are revealed by the equal area principle of a SG, due to their similarities in dynamical equations. Stability margin are evaluated by calculation of Critical Clearing Time (CCT), in which a small PLL bandwidth results in a large stability margin. The numerical method developed for CCT can be applied to study other parameters' effects on frequency stability, e.g. line impedance.
3) Under high voltage conditions, where the line resistances can be neglected, the reactive power ($Q$) controller has negligible effects on frequency stability compared to the active power ($P$) controller. Furthermore, by reducing the $P$ controller bandwidth, stability margin can be further improved.